# Qtier-*Rapor*: Managing Spreadsheet SYSTEMS & Improving Corporate Performance, Compliance and Governance.

### Keith Bishop

Qtier Software Limited
18 Queen Street, Market Drayton, Shropshire, TF9 1PX, UK
info@qtier.com

**ABSTRACT**

*Much of what EUSPRIG discusses is concerned with the integrity of individual spreadsheets. In businesses, interlocking spreadsheets are regularly used to fill functional gaps in core administrative systems. The growth and deployment of such integrated spreadsheet SYSTEMS raises the scale of issues to a whole new level.*

*The correct management of spreadsheet systems is necessary to ensure that the business achieves its goals of improved performance and good corporate governance, within the constraints of legislative compliance – poor management will deliver the opposite !*

*This paper is an anatomy of the real-life issues of the commercial use of spreadsheets in business, and demonstrates how Qtier-Rapor has been used to instil best practice in the use of integrated commercial spreadsheet systems.*

## 1 INTRODUCTION

Spreadsheet systems aren't designed – they grow organically …. as does the effort needed to control them !

A typical 'system' starts with the recognition of an inadequacy in the existing formal systems, and an individual builds a quick fix spreadsheet to bridge the gap. As time goes by, changes are made to make the system 'work better'; eventually the enhanced function comes to the attention of a manager, and the 'system' is released departmentally; and then company wide.

Users within the business have differing views on the function needed to support their own aspect of the business – they love the new system because they can change it to suit their own way of working.

Within a short space of time, there are a number of 'versions' of the spreadsheet system, all with common features….

- lack of documentation
- lack of ownership
- lack of support
- lack of change control

The improvements in contribution to corporate performance plateau – but the decline in performance, caused by an inability to support the monster that has been created, just continues on down.

This document follows the anatomy of a spreadsheet based budgeting system, and the use of Qtier-Rapor to arrest its decline, improve the overall performance of the business system and instil compliance features that encourage best practice and good governance.

## 2    CORPORATE UNDER-PERFORMANCE – THE DECLINE AND FALL OF A SPREADSHEET ?

The business is a dynamic, high growth operation with a heavy reliance on IT.  The core business application contains detailed financial budgets, but has limited function to assist in arriving at those budgets.

When the business was in its infancy a simple budget spreadsheet was created by the finance department.  All the essential failings were 'designed in' from the outset.
- fixed format for each department
- one sheet per department
- manual consolidation
- control owned by finance
- changed to suit the needs of the business, as and when required
- totally undocumented
- reliance on key individuals

With only a small number of departments, each owned by a director, filling in the departmental spreadsheet was simple.  With some adjustments, the budget numbers were manually loaded into the finance system.

The company diversified, so different departments had different needs. One fixed-format sheet per department meant that the format for every department grew to the biggest common denominator.

- The company grew – more departments meant more worksheets in the workbook.
- Departments grew in separate locations – the budget had to be deployed by email.
- The company matured – budgeting was no longer a view of the Profit & Loss, but needed to perform to certain key indicators (KPI's)
- The management structure matured – directors had departmental managers and cost centre owners.

Security became an issue - managers were not necessarily privy to the budgets of other departments or the business as a whole. Salary rises (or lack of them) were central to the budget process, but couldn't be on general display.

Departments continued to flex their budget numbers ready for the next budgeting round.  As a consequence, departments maintained multiple versions of their own budget.

Consolidation became an exponential problem – particularly in knowing which versions to consolidate, and their state of readiness.

Quarterly re-budgeting, to forecast the next quarter's and whole year's outcome on the basis of current actual performance, became a necessity for good governance.
The business was in acquisition and re-structuring mode, so scenarios were played out with the numbers.

The business ended up with:
- a 6Mb budget workbook, containing 30 departmental worksheets.
- populated by finance with the previously signed off budget
- emailed to every departmental manager for completion of just their elements
- returned by email to finance (30 copies)
- consolidated manually by finance
- reflected in a pivot table for management discussion
- de-constructed to email back out to departments for adjustment
- departments re-entered adjustments from their own 'next version'
- re-consolidated by finance for the next round of consideration
- up to five cycles to reach an approved budget
- end results entered into finance system for actual vs budget tracking
- repeated every quarter

No budgeting 'package' had the flexibility to cope with the business requirement, but the use of spreadsheets seemed to break every rule of best practice for corporate governance or compliance, and was certainly not improving corporate performance: there had to be a better way.

## 3   MANAGING SPREADSHEET SYSTEMS

Managing the validity of individual spreadsheets is core to control and is much discussed in EUSPRIG documents.

Managing individual spreadsheets in the context of Corporate Compliance, involves:
- Access control
- Change Control
- Version Management
- Functional audit with separation of roles

Managing Spreadsheet SYSTEMS adds extra dimensions of complexity:
- Interoperability between spreadsheets
- Dependency on prior events
- Sharing data
- Data structures
- Consolidation
- Automation of processes

The following discussion is framed around the requirements of a rigorous budgeting system that has been designed and built using Qtier-Rapor in order to address the improvement of corporate performance within a compliant framework.

### 3.1 Access Control

We need to distinguish between access to a spreadsheet and access to relevant data within that spreadsheet structure, based on role.  There is a difference between permission to access the structure of the spreadsheet, in order to carry out appropriate calculations, and permissions to access data relevant only to the user.

In the budget system, the user is authorised access to a spreadsheet template based on role. Each authorised user is maintained in a department table, so they will only see data relevant to their department(s). The data is stored centrally, and access is filtered by understanding the role and permissions of the user.

### 3.2 Change Control

In this context, we need to distinguish between a User, with permission to access and change data to which they are authorised, and an Editor, with permission to access and change structure & formulae.

Authorised users can enter their own budget data, but have no access to the structure of the budget. Every budget spreadsheet is secured against any change to locked (ie non-data entry) cells without specific editing authorisation.

Authorised Owners, Editors & Auditors must co-operate in order to change the structure of the system.
- Live spreadsheet copied to Work in Progress (WIP) by Owner.
- Changes made by an Editor and spreadsheet flagged for auditing.
- Auditor uses best practice audit tools to ensure integrity of structure.
- Auditor passes spreadsheet for release, or refers back to Editor for further changes.
- Owner archives current live version and releases new live version.

Every step of the process is automatically recorded in the audit log in order to produce change control documentation of who did what when. Segregation of Roles between Owner/Editor and Auditor, as specified in the corporate compliance policy, is enforced by Qtier-Rapor.

### 3.3 Version Management

This has two meanings in the context of our budget system. One is the actual version control of each spreadsheet, ensuring that there is only one live version of spreadsheet logic in use at any one time, and that previous versions are archived.

The second meaning is version control of the data content of the budget system, to ensure that each budget-round is in synchronisation and that departments can work on a WIP 'next version' of their budget whilst the current version is still under management consideration.

### 3.4 Interoperability and Dependency

A single spreadsheet can be used in isolation. Within a spreadsheet system we must introduce dependency checks because certain actions dictate the validity of further actions.

For example, consolidating the results of all departments is invalid if some of those departments have not yet completed their budgeting input.

A dependent process requires knowledge of the status of the factors on which it depends. An upstream process requires the ability to indicate its status.
The system requires a mechanism to pass that status between the processes.

In the budget system, the state of readiness of each element of the budget was designed into the process, and the data features of Qtier-Rapor were used to make the status available in real time.

Fig. 1  Consolidation Dependency - Status of Completion of Processes

### 3.5 Sharing Data & Data Structures

A single spreadsheet contains its own data, or can download data from an external link. A spreadsheet system requires that a spreadsheet can have access to data that has been generated by another spreadsheet in the system, as well as having access to data from other sources.

Microsoft allows the linking of spreadsheet cells to those of another spreadsheet, as well as the ability to access data from external sources, and this is the compliance weak point of any spreadsheet system. The Allied Irish Bank fraud was not caused by an error in a spreadsheet structure, it was caused by an error in the location from which external exchange rates were extracted.

A Microsoft spreadsheet can extract data from another spreadsheet in the chain provided that
- It has access to the current version of the spreadsheet.
- It can access the spreadsheet using the file name embedded - which means that, at some point, either a filename must be changed or all versions of the source of the data must have the same name !!
- The user has certain knowledge of the status of the spreadsheet providing the data, to know that the data is in a completed state.

In order to be flexible, Microsoft also allows the spreadsheet user the option of refreshing the data or not – effectively leaving it up to the user to decide if they wish to proceed on the basis of current data or old data !

It is not possible to dictate by 'Corporate Policy' that that the data being transferred between two spreadsheets is current and ready, and that the process cannot proceed until the data has been transferred.

IT systems achieve this every day by the use of databases. Spreadsheets can be integrated to databases in order to extract information, and this is a far better methodology than retaining shared data in a spreadsheet. A spreadsheet system based only on downloads would be restricted to use only as a reporting tool – to build a secure system, the spreadsheets need to be able to securely write information back to a database.

Qtier-Rapor can control data transfer both from and to a structured database, and this is the way that it retains, consolidates and disseminates data within a system; including metadata about corporate controls, access permissions and the status of completion and dependency. This gives true 'best practice' separation of data and formulae.

The Qtier budgeting system uses a 'SQL Server' database, but it could use anything from 'Access' to 'Oracle'. Key to the design of a system is that the data structure can be designed using the same techniques as for an IT system, and that the data can be shared with an IT system. Proper key structures prevent duplication of records; good structure design minimises storage and maximises performance.

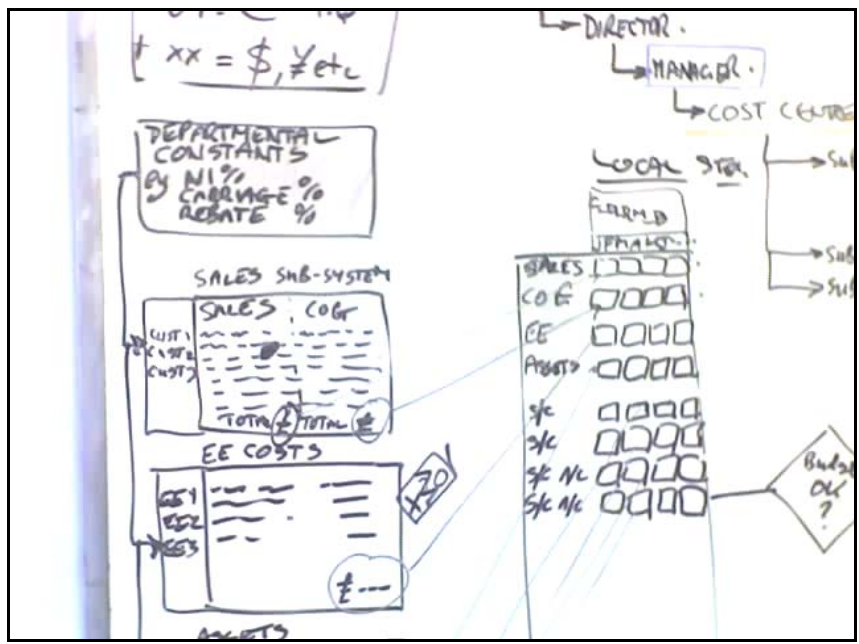

**Fig. 2 Whiteboard of process & database design**

Data required for the Qtier budgeting process is extracted from the database in the full knowledge of the identity of the user. Since the data includes the departmental access constraints for the user, then the process can dictate …

- That only that budget data to which the user has authority is presented, but presented within an audited template structure that is common across all departments.
- The process can only proceed with the current data direct from the database, and the user has no ability to interfere with that data being loaded.
- Part of the data relates to the status of completion of a previous process.
- Part of the process relates to its dependency on that previous process.
- The process can itself write system data and status data back to the database for use by other processes.

In the case of the budgeting system, we access a corporate data warehouse for historic actual finance numbers, and maintain a set of Qtier-Rapor budget files within the data warehouse so that they can be accessed by every authorised budget user and can also be accessed by the corporate system for import into the core financials.

### 3.6 Consolidation

The previous spreadsheet based system was consolidated from 30 different spreadsheets in order to prepare a management view within a pivot table.

This required…

- That all of the spreadsheets were returned to a central 'consolidator' by email.
- That the consolidator copied the data into his own 30 worksheet structure.
- That the consolidator therefore also copied any formula changes into the central structure.
- The consolidator resolves any structure differences included by the department.
- The business had to wait until all spreadsheets had been returned, and had no knowledge of their status without phoning each departmental head.

By separating out the budgeting data from the structure of the spreadsheets …

- Each department had access to only those budget sections relevant to them and consequently only needed to enter data for a smaller sub-set of relevant headings.
- The state of readiness (not started / work in progress / completed) was immediately visible to senior management for every section of every department.
- Consolidation direct from the database was immediately available once all data entry was completed.
- Individual departments could begin work on refining the budget whilst the current version was under consideration, giving significant business speed improvements to the whole process.
- Senior managers were able to see consolidations of their departments during each budgeting round, regardless of departmental completion, so that they could guide the process.
- By including data from previous versions of the budget, from previous year's budgets and from actual values in the finance systems, a series of management KPIs and comparison graphs were immediately available in consolidation or detail.

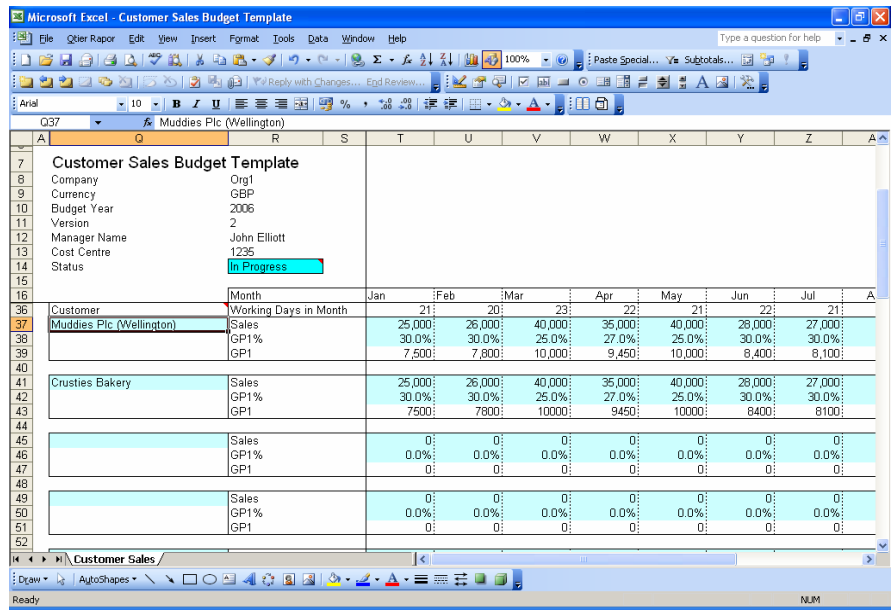

**Fig. 3 Specific Cost Centre data Entry**

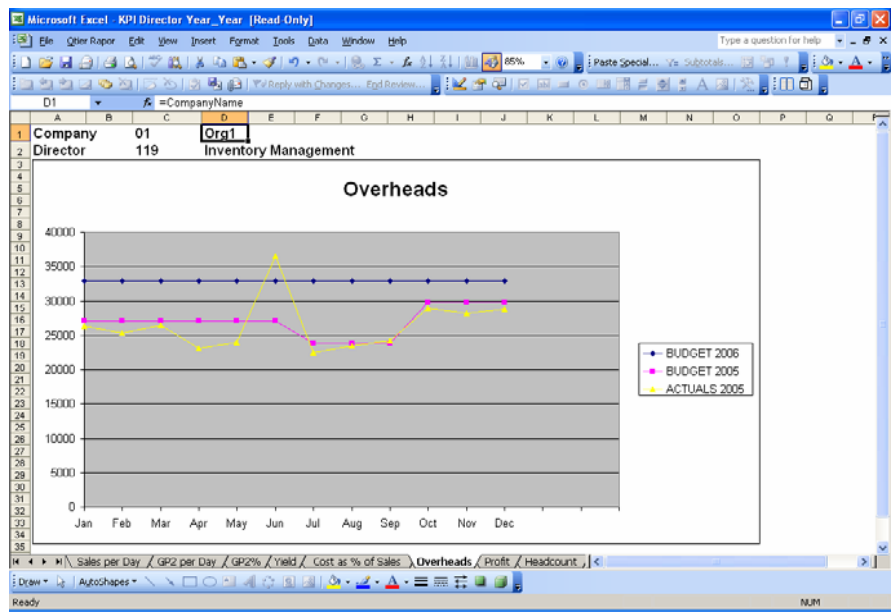

**Fig. 4 KPIs Incorporating Numbers from Data Warehouse**

### 3.7 Automation

Good corporate governance relies on being able to mandate policies that systems will follow. Governance also relies on feedback that the policies are being actioned as mandated.

This means that good governance should require that the mandated policies are implemented automatically, to remove any human discretion, and that the feedback is automated to remove any editing by those to whom it refers.

| Policy | Qtier Automated Action |
|---|---|
| Access Control | Access to processes based on permissions by user role. Access to data based on user department and role. |
| Change Control | Complete change process from WIP to Live, with every action logged. Management and archiving of structure versions. Capability to version control entered data. |
| Segregation of Roles | Where mandated, it is not possible for the editor of a spreadsheet to audit it; and not possible for an auditor to release an audited spreadsheet into the live environment. |
| Security of Data | All critical data held in a structured database on a secure server. Full adherence to corporate server data security & back-up policies. |
| Integrity of Data | All data held in a structured database. All access to data filtered on user permissions. Status information about the readiness of the data stored in the database. Dependency checking between processes. No user discretion on the use or location of the data. Authentication stamps on screen displays and printouts. |
| Audit visibility | Templates audited and locked down. Automated logging in a secure Audit database. Visibility of tampering attempts. Time-stamped trail of every usage. Time-stamped trail of every change. Visibility of policy violation attempts. |
| Automation of controls | No user interference with the source, location or transfer of data between processes. No user ability to circumvent audit logging. Replacement of macros with secure scripts in critical spreadsheets. Automation of tasks to enable preparation and distribution of secure Board reporting documents without user intervention. |

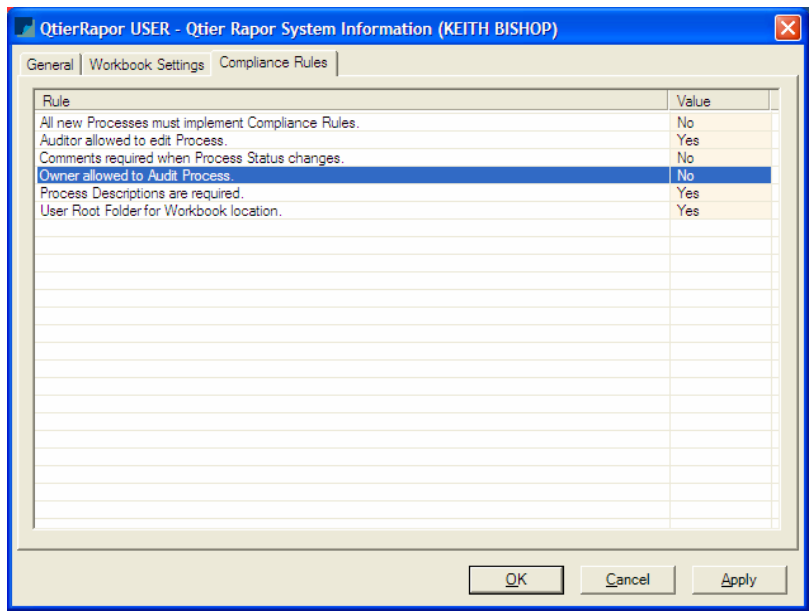

**Fig. 5  Example of Compliance Rules**

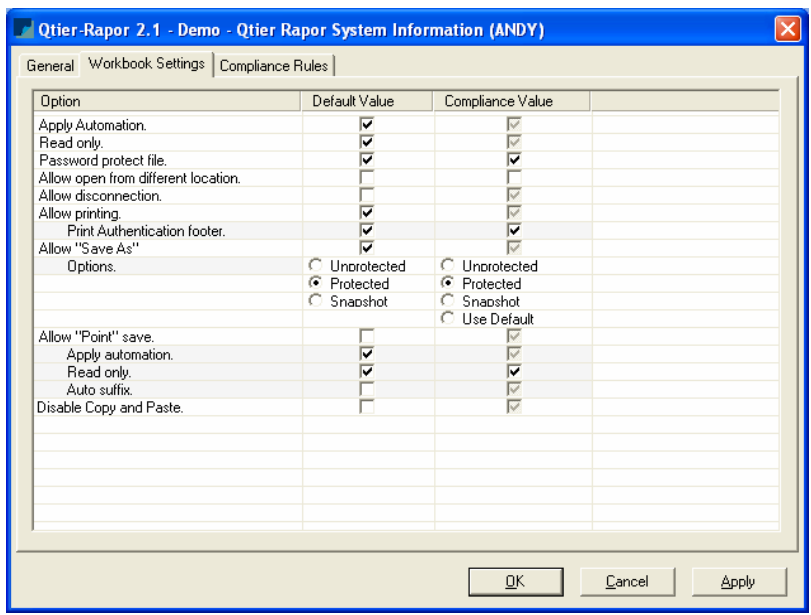

**Fig.  6  Example of Compliance Implementation - Default Settings**

## 4    SUMMARY

*The correct management of spreadsheet systems is necessary to ensure that the business achieves its goals of improved performance and good corporate governance, within the constraints of legislative compliance – poor management will deliver the opposite !*

This paper follows the specific 'spreadsheet systems' issues raised in a real life example. and demonstrates how software technology can be used to resolve them.

By re-engineering their budgeting system as described, and consequently being able to mandate and manage policies in line with best practice corporate governance, the business significantly improved its corporate performance by delivering a budget within 25% of the previous cycle time, whilst ensuring that compliance with good governance was higher than it had ever been.

Subsequent business re-structuring was accomplished with simple additions to the company/department/cost centre tables, rather than wholesale spreadsheet changes.

The total time spent to undergo training in the Qtier-Rapor system and to create the budgeting system was 10 days for a team of two.  This was recouped in time savings on the first live run.

*Business speed with 'best practice' compliance.*

## 5    FUTURE PERSPECTIVE

Businesses will continue to need the flexibility of spreadsheets, but increasingly find that some of that flexibility becomes a barrier to achievement of legislative compliance and best practice governance.  This is particularly so where spreadsheets are integrated with each other to build a system.

Qtier will continue to enhance the usability and power of functions, in order to ensure that design of spreadsheet systems is undertaken within the Qtier-Rapor framework as a matter of choice rather than mandate.

Blank Page